\documentstyle[multicol,aps,prb,epsfig]{revtex}
\input epsf
\begin{document}
\draft \preprint{Revised Version. Submitted to Phys. Rev. B }

\title{ A simple model for dynamical melting of moving vortex \\
lattices interacting with periodic pinning}

\author{ Cl\'ecio C. de Souza Silva}
\address{Departamento de F\'{\i}sica, Universidade Federal de Pernambuco
50670-901, Recife-PE,  Brasil }
\author{Gilson Carneiro$^{*}$}
\address{Instituto de F\'{\i}sica, Universidade Federal do Rio de Janeiro,
C.P. 68528, 21945-970, Rio de Janeiro-RJ, Brasil }
\date{\today}
\maketitle
\begin{abstract}

Melting of moving vortex lattices in clean superconducting films with periodic
pinning is studied by a dynamical ``cage" model, based on a mean-field
treatment of Langevin's equations for the whole vortex lattice, assuming
elastic flow. In the frame moving with the velocity of the vortex lattice
center of mass, the model  describes forced vibrations of a single vortex tied
to a spring. The  vortex displacements due to thermal fluctuations and to the
periodic pinning force, and the relationship between the driving force and the
vortex velocity (V-I curves) are obtained by a perturbation method, valid for
high velocities and for both weak and strong periodic pinning. The dynamical
melting temperature is calculated as a function of the vortex velocity using
Lindemann's criterion. Application to a square defect lattice gives, for strong
pinning, dynamical melting lines and anisotropic V-I curves that agree
qualitatively with numerical  results.

\end{abstract}

\begin{multicols}{2}
\narrowtext

\section{introduction}
\label{sec.int}

The study of dynamical phases of lattice structures moving in a periodic medium
has received a great deal of attention lately due to its relevance to  physical
systems such as adsorbed atomic layers in boundary lubrication \cite{slfr},
vortices in Josephson Junction Arrays (JJA)\cite{jja} and  in films with
artificial defect lattices\cite{ppa}.

Numerical studies of driven vortices in clean superconducting films interacting
with a periodic lattice of columnar defects (CD)\cite{gmc,gmcr} and in JJA
\cite{dom1,dom2} find dynamical melting of a moving vortex-lattice (VL) into a
moving vortex liquid. In these systems  the moving vortices order in a VL at
low temperatures  when the  driving force magnitude is sufficiently large.
Melting of the VL is found to take place at a temperature that, for a given
vortex density, decreases with the  driving force magnitude, approaching the
equilibrium VL melting temperature when this magnitude approaches infinity.
These studies  also show that the vortices center of mass (CM) velocity is in
general not parallel to the driving force. This leads to anisotropic V-I
curves, with voltages both longitudinal and transverse to the direction of the
applied current.

In this paper we show that dynamical melting lines and  anisotropic V-I curves
similar to those found in the above mentioned numerical studies can be obtained
using a simple  model for the vortex dynamics.

In driven vortex systems interacting with random or periodic pinning, moving
vortex lattices are expected to occur at low temperatures and in the limit of
very large CM velocities, because the pinning potential is averaged in the
direction of motion, as first pointed out by Schmid and  Hauger \cite{hdp}. For
periodic pinning, Ref.\onlinecite{gmc} shows that in this limit the moving
vortex lattice  can be commensurate or incommensurate with the periodic pinning
potential, depending on the direction of motion and on the vortex density.
Commensurate lattices occur only for motion along directions of high symmetry
of the pinning potential, such as [1,0], [0,1], [1,1] and [-1,1] for square
symmetry, if the vortex density is not too high \cite{gmc,rof,rz}. For motion
along other directions, the moving vortex lattices are incommensurate.  Recent
numerical simulations find that both  moving commensurate and incommensurate
lattices may undergo dynamical melting into a moving
liquid\cite{gmc,gmcr,dom1,dom2,rz}. However, dynamical melting of commensurate
lattices is much more complicated than that of the incommensurate ones. The
reason is that the moving commensurate lattices are pinned in channels running
along the direction of motion (transverse pinning), and melting into a liquid
occurs simultaneously with depinning from the channels\cite{gmc,gmcr,rz}. In
this paper we consider only dynamical melting of incommensurate lattices.

Dynamical melting of vortex lattices is still poorly understood. It is unclear
from the simulation results reported so far if the change from lattice to
liquid spatial symmetries is a true phase transition, with accompanying
singularities in correlation functions, or merely a crossover.  In this paper
we do not offer any insight into this fundamental question. Our aim is to
estimate the dynamic melting temperature based on a well know method for the
equilibrium melting temperature, namely Lindemann's criterion. This method was
proposed in a classical paper by Houghton, Pelcovitz and Sudb\o \cite{hps}.
They estimate the equilibrium VL melting temperature for clean bulk
superconductors  by  applying  Lindemann's criterion to the mean-square
vortex-line displacement caused by thermal fluctuations. A simple model to
estimate the same temperature was proposed later by Frey, Nelson and Fisher
\cite{fnf}. It considers a single vortex line trapped in a harmonic potential
representing its interaction with the other vortex lines in the lattice,
assumed straight and fixed at their equilibrium lattice positions. We refer to
this model here as the equilibrium cage model. The equilibrium melting
temperature is estimated by applying Lindemann's criterion to the mean-square
thermal fluctuations of the caged vortex line. The result agrees with that of
Ref.\onlinecite{hps}, if the harmonic potential  is properly chosen\cite{fnf}.
This simple idea can also be applied to two-dimensional VL melting in thin
films. In this case  the equilibrium cage model consists of a point vortex tied
to a spring, representing its interaction with the other point vortices,
assumed fixed at their lattice positions. The melting temperature  is obtained
by applying Lindemann's criterion to the caged vortex mean square displacement
due to  thermal fluctuations  \cite{fnf}. The resulting melting temperature
agrees  with the one obtained from dislocation unbinding theory\cite{fish2}, if
the spring constant is properly chosen. Our model, which we call the dynamical
cage model, generalizes the two-dimensional equilibrium one for a moving vortex
lattice. The dynamical melting temperature is estimated by applying Lindemann's
criterion to the mean-square vortex displacements caused by thermal
fluctuations and by motion in the periodic pinning potential. Application of
Lindemann's criterion to estimate the dynamical melting line of vortices in
bulk superconductors interacting with random pinning was first proposed by
Koshelev and Vinokur\cite{kos}.

We start from Langevin's equations describing two-dimensional vortices at
temperature $T$, interacting between themselves and with a periodic pinning
potential, and driven by an external force. We consider vortex motion at large
CM velocities and make the following assumptions: i)  The moving vortices order
in a lattice, incommensurate with the periodic pinning potential. ii) The
vortex lattice flows elastically. This means that there are no dislocations and
that vortex-vortex interactions can be approximated by harmonic forces.  We
then propose a mean-field like treatment of the resulting equations of motion.
This approximation reduces the problem to that of a single vortex tied to a
spring, both moving with the CM velocity, and interacting with the periodic
pinning potential. We find that the dynamical cage model is governed by two
equations of motion. One for the vortex displacement in the frame moving with
the vortex lattice CM velocity (CM frame), and another relating the driving
force and the  CM velocity. The vortex displacement equation is  found to be
nonlinear, unlike that for the equilibrium cage model, due to the
vortex-pinning  interaction. These equations  are solved by a perturbation
method, similar to that introduced by Schmid and Hauger \cite{hdp} for weak
random pinning forces. It consists in expanding the time-dependent periodic
pinning force (in the CM frame) in powers of the vortex displacement. We find
that the perturbation expansion is valid at large CM velocities for both weak
and strong pinning. From these solutions we obtain  expressions for the vortex
mean square displacement and for the relationship between the CM velocity and
the driving force. The  dynamical melting temperature is obtained by applying
Lindemann's criterion to the caged vortex mean-square displacement.
  These results are applied to a square
pinning array, and  compared to numerical simulation ones.

This paper is organized as follows. In Sec.\ \ref{sec.dyncm} we derive the
dynamical cage model equations of motion and  their perturbation theory
solution.  In Sec.\ \ref{sec.sqlat}. we apply these results  to strong pinning
by a  square array of pinning centers.  The conclusions of the paper are
presented in Sec.\ \ref{sec.concl}. Details of the perturbation calculation are
given in the Appendix.

\section{Dynamical Cage Model}
\label{sec.dyncm}

We consider $N_v$ two-dimensional vortices at temperature $T$ interacting
between themselves and with a periodic  potential produced by an array of
pinning centers, and driven by a force ${\bf f}_{d}$. The equations of motion
are \cite{ehb}
\begin{equation}
\eta \frac{d{\bf r}_j}{dt}= {\bf f}_{d} + {\bf F}^{\text{\it v-v}}_j + {\bf
F}^{\text{\it v-p}}({\bf r}_j) + {\bf \Gamma}_j\;, \label{eq.lan}
\end{equation}
where $j=1,...,N_v$, $\eta$ is the friction coefficient,
\begin{equation}
{\bf F}^{\text{\it v-v}}_j=-\sum^{N_v}_{i\neq j=1} {\bf \nabla}_jU^{\text{\it
v-v}}({\bf r}_j-{\bf r}_i)\; , \label{eq.fvv}
\end{equation}
is the force of interaction with other vortices, $U^{\text{\it v-v}}({\bf r})$
being the vortex-vortex interaction potential in two dimensions,
\begin{equation}
{\bf F}^{\text{\it v-p}}({\bf r}_j)= \sum_{\bf Q}(-i{\bf Q}) U^{\text{\it
v-p}}_{\bf Q} e^{i{\bf Q}\cdot {\bf r}_j}\;, \label{eq.fvpl}
\end{equation}
is the force of interaction with the periodic pinning array, ${\bf Q}$ denotes
the pinning array reciprocal lattice vectors, and $U^{\text{\it v-p}}_{\bf Q}$
is the Fourier transform  of the vortex-single pin interaction potential, and
${\bf \Gamma}_j$ is the random force, satisfying
\begin{equation}
\langle \Gamma_{j\alpha}(t)\Gamma_{l\beta}(t^{\prime})\rangle =
2k_BT\eta\delta_{j,l} \delta_{\alpha,\beta}\delta(t - t^{\prime}) \; ,
\label{eq.gam}
\end{equation}
where $\langle\cdots\rangle$ denotes average over the random force
distribution.

We assume that the vortices center of mass (CM) moves with constant velocity
${\bf v}$, defined, as usual, by
\begin{equation}
{\bf v}= \frac{1}{N_v}\sum^{N_v}_{j=1} \frac{d{\bf r}_j(t)}{dt}\; .
\label{eq.vcmd}
\end{equation}
Using Eq.\ (\ref{eq.lan}), we find that
\begin{equation}
\eta{\bf v} = {\bf f}_d +\frac{1}{N_v}\sum^{N_v}_{j=1} {\bf F}^{\text{\it
v-p}}({\bf r}_j(t))\;. \label{eq.vcmfd}
\end{equation}
To derive Eq.\ (\ref{eq.vcmfd}) we use the fact that for large $N_v$, the
random force term is negligible, since $\sum^{N_v}_{i=1}{\bf \Gamma}_j \sim
\sqrt{N_v}$. According to Eq.\ (\ref{eq.vcmfd}), in order that ${\bf v}$ is
constant ${\bf f}_d$ must depend on time as well as on the random force. To
obtain physical results we average Eq.\ (\ref{eq.vcmfd}) over time and over the
random force distribution, that is
\begin{equation}
 \eta{\bf v} = {\bf F}_d
+\frac{1}{N_v}\sum^{N_v}_{i=1}\frac{1}{\tau}\int^{\tau}_0 \big\langle{\bf
F}^{\text{\it v-p}}\big({\bf r}_j(t)\big)\big\rangle \,, \label{eq.vcm}
\end{equation}
where ${\bf F}_d$, a constant vector, denotes the average of ${\bf f}_d$ over
time and over the random force distribution and $\tau$ is a time large compared
with the characteristic times in Eq.\ (\ref{eq.lan}). We interpret ${\bf F}_d$
as the force due to the applied current, so that Eqs.\ (\ref{eq.vcm}) give the
V-I curves.

Now we consider a  vortex-lattice flowing elastically and transform to the CM
frame. We write the vortex positions as
\begin{equation}
{\bf r}_j(t)={\bf R}_j+ {\bf v}t + {\bf u}_j(t)\; , \label{eq.elfl}
\end{equation}
where  ${\bf R}_j$  and  ${\bf u}_j(t)$ ($j=1,...,N_v$) are, respectively, the
vortex equilibrium positions and displacements from equilibrium in the CM
frame. In elastic flow the displacements ${\bf u}_j(t)$ are small and the
vortex-vortex interactions can be approximated by harmonic forces. Our
dynamical cage model considers a particular vortex, and substitute its elastic
interaction with the other vortices in the lattice by a spring of constant
$\kappa$. The equation  of motion for this vortex in the CM frame is found to
be
\begin{eqnarray}
\eta \frac{d{\bf u}_l(t)}{dt} &=& -\kappa {\bf u}_l(t) + {\bf F}^{\text{\it
v-p}}({\bf R}_l + {\bf u}_l(t)+ {\bf v}t)
\nonumber \\
& & + {\bf f}_{d} - \eta {\bf v} + {\bf \Gamma}_l\;. \label{eq.cag1}
\end{eqnarray}
The  relationship between the spring constant  and the vortex-vortex
interaction potential will be presented shortly. Our dynamical cage model
approach  is similar to the mean-field theory introduced by Fisher\cite{fish}
for sliding charge-density waves interacting with random pinning.

Our objective here is to use Eq.\ (\ref{eq.cag1}), together with Eq.\
(\ref{eq.vcm}), to calculate the vortex mean-square displacement required for
Lindemann's criterion, and to obtain the ${\bf v}$ vs. ${\bf F}_d$ relationship
(V-I curves). First we note that  Eq.\ (\ref{eq.cag1}) depends on the
particular vortex $l$, through the ${\bf F}^{\text{\it v-p}}$- term. Since
${\bf F}^{\text{\it v-p}}$ is periodic in the pinning-array lattice, its
argument in  Eq.\ (\ref{eq.cag1}) can be reduced, at any instant $t$, to a
position within the pinning-array  primitive unit cell. As mentioned  in Sec.\
\ref{sec.int}, we consider only  moving incommensurate vortex lattices. In this
case the reduced vortex positions are uniformly distributed over the unit cell
area. Accordingly, we define the vortex mean square displacement as
\begin{equation}
u^2 = \frac{1}{N_v}\sum^{N_v}_{j=1}\frac{1}{\tau}\int^{\tau}_0
\big\langle\mid{\bf u}_j(t)\mid^2\big\rangle\;. \label{eq.usq}
\end{equation}

According to the above considerations, our dynamical cage model is described by
the set of equations
\begin{eqnarray}
\eta \frac{d{\bf u}_l(t)}{dt} &=& -\kappa {\bf u}_l(t) + {\bf F}^{\text{\it
v-p}}({\bf R}_l + {\bf u}_l(t)+ {\bf v}t)
\nonumber \\
& & -\frac{1}{N_v}\sum^{N_v}_{j=1} {\bf F}^{\text{\it v-p}}({\bf R}_j + {\bf
u}_j(t)+ {\bf v}t) + {\bf \Gamma}_l\,, \label{eq.cag2}
\end{eqnarray}
and
\begin{equation}
\eta{\bf v} = {\bf F}_d
+\frac{1}{N_v}\sum^{N_v}_{j=1}\frac{1}{\tau}\int^{\tau}_0 \big\langle{\bf
F}^{\text{\it v-p}}({\bf R}_j + {\bf u}_{\bf v}t)\big\rangle \,.
\label{eq.vcm2}
\end{equation}

To  obtain $ {\bf u}_l(t)$ we solve  Eq.\ (\ref{eq.cag2}) by  the perturbation
theory method introduced in Ref.\onlinecite{hdp}, which consists in expanding
${\bf F}^{\text{\it v-p}}$ in powers of ${\bf u}_i(t)$ and solving the
resulting equations by iteration. We keep only terms  to first order. The
relationship between ${\bf v}$ and ${\bf F}_d$ are obtained from Eq.\
(\ref{eq.vcm2}) by a similar expansion.  The details of these calculation are
given in the Appendix.

For the mean square displacement we obtain to leading order
\begin{eqnarray}
u^2 &=& \frac{2k_BT}{\kappa} + \sum_{\bf Q} \frac{Q^2|U_{\bf Q}|^2}
{\eta^2({\bf Q}\cdot{\bf v})^2+\kappa^2}
\nonumber \\
& & + \frac{2k_BT}{\kappa}\sum_{\bf Q}\frac{Q^4 |U_{\bf Q}|^2} {\eta^2({\bf
Q}\cdot{\bf v})^2+4\kappa^2} \; . \label{eq.usq2}
\end{eqnarray}

To the  same order of approximation the ${\bf v}$ vs. ${\bf F}_d$ relationship
is
\begin{eqnarray}
   {\bf v} &=& \frac{{\bf F}_d}{\eta} -
   \sum_{\bf Q}\frac{{\bf Q}\:Q^2|U_{\bf Q}|^2}
   {\eta^2({\bf Q\cdot v})^2+\kappa^2}\,{\bf Q\cdot v}
   \nonumber \\
& &-\frac{k_B T}{\kappa}\sum_{\bf Q}\frac{{\bf Q}\:Q^4|U_{\bf Q}|^2}
   {\eta^2({\bf Q\!\cdot v})^2+4\kappa^2}{\bf Q\cdot v} \;.
   \label{eq.vcm3}
\end{eqnarray}

According to Lindemann's criterion the VL melts  when $u^2 = c_L^2 a_v^2$,
where $a_v$ is the VL lattice parameter. Thus, our dynamical cage model only
makes sense  as long as $u^2 < c_L^2 a_v^2$. This requires that  in  Eq.\
(\ref{eq.usq2}) both $k_BT/\kappa$, and  the velocity dependent terms are
sufficiently small compared to $a^2_v$. The latter are so for large enough $v$.
To state more precisely the conditions under which the perturbation expansion
in Eq.\ (\ref{eq.usq2}) is valid, we restrict our considerations to periodic
pinning forces that can be represented by Eq.\ (\ref{eq.fvpl}) with only a few
non-zero $U_{\bf Q}$. Typically those for ${\bf Q}$ corresponding to nearest
and next-nearest neighbor reciprocal lattice points. In these cases $Q\sim
2\pi/a_p$, $a_p$ being the pinning array lattice parameter. We also assume that
the number of vortices is comparable to the number of pins,  so that $a_v \sim
a_p$. Under these assumptions, the condition for the smallness of both velocity
dependent terms in  Eq.\ (\ref{eq.usq2}) depends  on the relative strengths of
the pinning and elastic forces. For weak pinning, defined as  $|U_{\bf Q}|\ll
\kappa a^2_p/2\pi$, these terms are always small, independent of ${\bf v}$, so
that the effects of vortex

\begin{figure}[here]
\centerline{\includegraphics[scale=0.33]{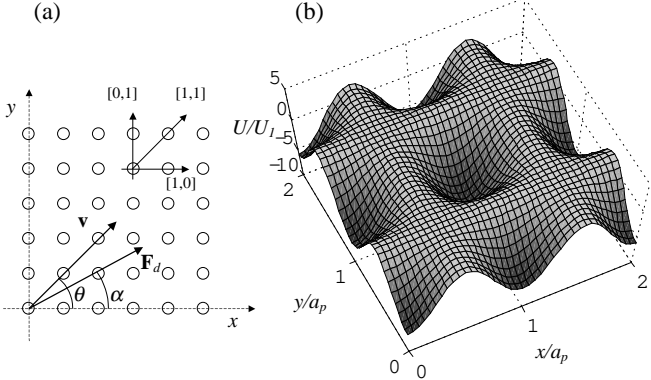}}
\caption{(a) Pinning lattice, coordinate system and angles defining the
directions of the center of mass velocity, ${\bf v}$, and the driving force,
${\bf F}_d$. (b) Square pinning potential (Eq. (\ref{eq.pinpot})) with
$U_2=U_1/2$.} \label{fig.pin}
\end{figure}

motion on $u^2$ are negligible. For strong pinning,
that is $|U_{\bf Q}|\sim \kappa a^2_p/2\pi$, the velocity dependent terms in
Eq.\ (\ref{eq.usq2}) are small for all ${\bf Q}$, if $\;\eta v
\cos{\vartheta_{\bf Q}}\gg \kappa a_p/2\pi \;$, where $\vartheta_{\bf Q}$ is
the angle between the particular ${\bf Q}$ and ${\bf v}$. Similar
considerations apply to the perturbation expansion for ${\bf v}$, Eq.\
(\ref{eq.vcm3}). For strong pinning, the perturbation expansion  breaks down
if, for some ${\bf Q}$ entering the sums in Eq.\ (\ref{eq.usq2}),  ${\bf Q\cdot
v}=0$ ($\cos{\vartheta_{\bf Q}}=0$). In this case the velocity dependent terms
in Eq.\ (\ref{eq.usq2}) are not small. This breakdown can also be seen from
Eq.\ (\ref{eq.cag1}). If ${\bf Q}\cdot{\bf v}=0$ the pining force ${\bf
F}^{\text{\it v-p}}({\bf R}_l + {\bf u}_l(t)+ {\bf v}t)$  has a static
component, periodic in the direction perpendicular to ${\bf v}$, which gives
rise to a static vortex displacement comparable to $a_v$ for strong pinning.
Such a displacement indicates that our assumption that the  moving VL is
incommensurate with the pinning array lattice breaks down for motion along  the
directions for which ${\bf Q\cdot v}=0$.  There  are only a few directions of
motion for which ${\bf Q\cdot v}=0$ since, as mentioned above, only a few ${\bf
Q}$ enter the sums in Eq.\ (\ref{eq.usq2}). We  note that the terms in sum in
Eq.\ (\ref{eq.vcm3}) with ${\bf Q}\cdot{\bf v}=0$ do not contribute to ${\bf
v}$, so that there is no breakdown in the perturbation expansion for this
quantity.

In the limit of very large $v$ ($v \rightarrow \infty$) our dynamical cage
model reduces to the static one, since vortex motion averages out the pinning
potential (we assume ${\bf Q\cdot v}\neq 0$). In this limit $u^2$ reduces to
the equilibrium mean square displacement of a vortex tied to a spring, $u^2 =
\frac{2k_BT}{\kappa}$. Application of Lindemann's criterion gives for the
melting temperature $k_BT_m= \kappa a^2_v c^2_L/2$. To compare this  estimate
of $T_m$  with the dislocation unbinding theory one, $\kappa$ is identified
with $\kappa = (\partial^2 U^{v-v}(r)/\partial r^2)_{r=a_v}$, where
$U^{v-v}(r)=(\phi^2_0/8\pi \Lambda)\ln(r/\xi)$  is the  2D vortex-vortex
interaction potential\cite{fnf} ($\phi_0$ is the flux quantum and $\Lambda$ the
film effective penetration depth ). The result is $\kappa = \phi^2_0/8\pi
\Lambda a^2_v$ and $k_BT_m= c^2_L \phi^2_0/16\pi \Lambda$.  This melting
temperature  agrees with the dislocation unbinding theory one \cite{fish2} if
$c_L\simeq 0.12$. From here on we adopt for $\kappa$ and $c_L$ the above
mentioned values.

Application of  Lindemann's criterion to Eq.\ (\ref{eq.usq2}) gives the
velocity dependent melting temperature $T_{dm}({\bf v})$ as
\begin{eqnarray}
k_BT_{dm}({\bf v}) =\, \frac{  \frac{\kappa a_v^2 c_L^2}{2}  - k_BT_{pp}({\bf
v})} { 1 + \sum_{\bf Q} \frac{Q^4 |U_{\bf Q}|^2} {\eta^2({\bf Q}\cdot{\bf
v})^2+4\kappa^2}}\; , \label{eq.tm}
\end{eqnarray}
where
\begin{equation}
k_BT_{pp}({\bf v}) \equiv \frac{\kappa}{2}\sum_{\bf Q} \frac{Q^2|U_{\bf Q}|^2}
{\eta^2({\bf Q}\cdot{\bf v})^2+\kappa^2} \label{eq.tsh}
\end{equation}
is identified with an effective temperature, resulting from vortex vibrations
due to the periodic pinning force, that adds to the thermodynamic one in the
expression for $u^2$. The  temperature $T_{pp}$ is similar to the 'shaking
temperature' introduced by Koshelev and Vinokur \cite{kos} for random pinning.
It arises from a $T$-independent contribution to $u^2$, the second term in the
right hand side of Eq.\ (\ref{eq.usq2}).  There is also a $T$-dependent
contribution to $u^2$ in Eq.\ (\ref{eq.usq2}) that cannot be identified with an
effective temperature. However, this term leads to a small correction in
$T_{dm}({\bf v})$, since the denominator in Eq.\ (\ref{eq.tm}) is close to one.
Thus, the dynamical melting condition is essentially that $k_B(T_{dm}({\bf v})+
T_{pp}({\bf v}))= \kappa a_v^2 c_L^2/2$. The ${\bf v}$ dependence of
$T_{dm}({\bf v})$, Eq.\ (\ref{eq.tm}), is such that $T_{dm}({\bf v})$ decreases
with decreasing $v$.

The V-I curves follow from the relationship between ${\bf v}$ and ${\bf F}_d$,
Eq.\ (\ref{eq.vcm3}). In the  $v \rightarrow \infty$ limit ${\bf v}={\bf
F}_d/\eta$. For finite $v$, vortex motion is no longer along the drive
direction, leading to anisotropic V-I curves. The  ${\bf v}$ vs. ${\bf F}_d$
relationship also allows us to obtain the melting temperature as a function of
the driving force, which is more usual.

In the next section we study in detail the  dynamical cage model predictions by
applying them to a typical periodic pinning potential and carrying out
numerical calculations.

\section{ square pinning-lattice}
\label{sec.sqlat}

We consider a square pinning array with lattice parameter $a_p$, and assume
that the vortex-pinning interaction force is given by Eq.\ (\ref{eq.fvpl}) with
the $U^{\text{\it v-p}}_{\bf Q}$ chosen as
\begin{eqnarray}
U^{\text{\it v-p}}_{\bf Q} =
   \left\{
   \begin{array}{ll}
     U_1  & \text{for ${\bf Q}=\pm\frac{2\pi}{a_p}\hat{x}$,
            $\pm\frac{2\pi}{a_p}\hat{y}$;} \vspace{1mm}\\
     U_2  & \text{for ${\bf Q}=\pm\frac{2\pi}{a_p}(\hat{x}\pm\hat{y})$;}
            \vspace{1mm}\\
     0    & \text{otherwise.}
   \end{array}
   \right.
   \label{eq.pinpot}
\end{eqnarray}
The $x$ and $y$ axis are along the pinning lattice $[1,0]$, $[0,1]$ directions,
respectively, as shown in   Fig. \ref{fig.pin}(a). We choose $U_2=U_1/2$. This
gives the pinning potential shown in Fig. \ref{fig.pin}(b). We set $U_1/\kappa
a^2_p= 1/2\pi=0.16$. As discussed in Sec.\ \ref{sec.dyncm}, this value
corresponds to strong pinning.

\begin{figure}
\centerline{\includegraphics[scale=0.46]{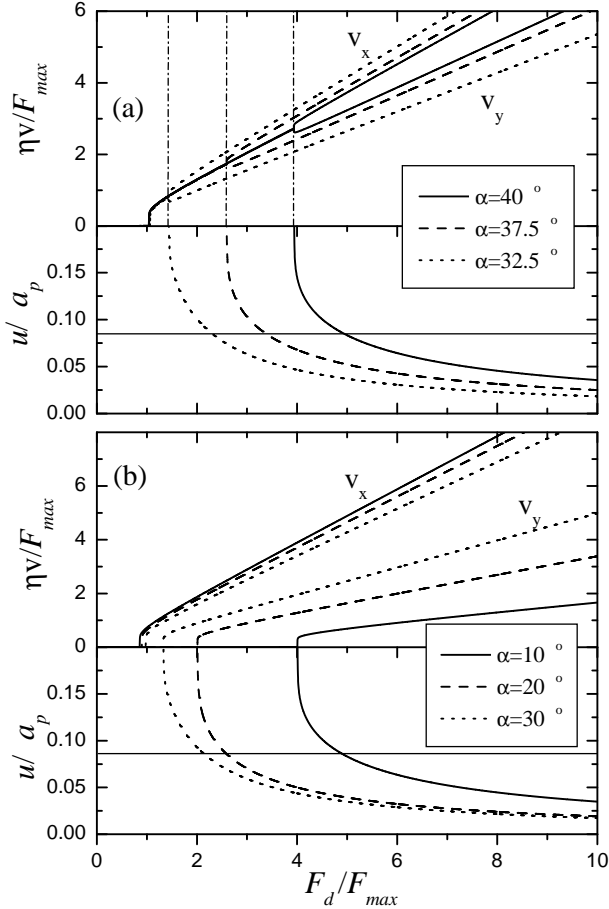}} \vspace{5mm}
\caption{Center of mass velocity components, $v_x$ and $v_y$, and
root-mean-square caged vortex fluctuations, $u$, vs. driving force magnitude,
$F_d$, for several driving force directions $\alpha$ at $T=0$. (a)
$32.5^o\leq\alpha\leq40^o$: vortex motion is trapped in the [1,1] direction
($v_x =v_y$) at the $F_d$ values indicated by the vertical dot-dashed lines in
the top panel. (b) $0<\alpha<32.5$: vortex motion is trapped in the [1,0]
direction ($v_y=0$). The horizontal lines in the bottom panels of (a) and (b)
are $u=c_La_v$ for $B=2B_{\phi}$ and $c_L=0.12$ (see text), where the VL melts.
} \label{fig.VIu}
\end{figure}

\begin{figure}[b]
\centerline{\includegraphics[scale=0.8]{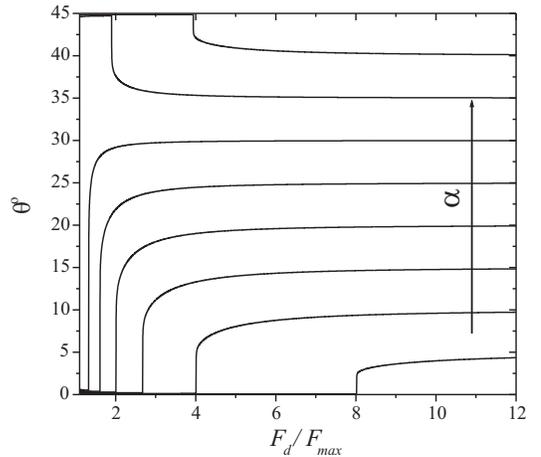}} \vspace{5mm}
\caption{Direction of motion, $\theta$, as a function of the driving force
magnitude for drive directions  in the  range $5^o \leq \alpha \leq 40^o$ in
regular increments of $5^o$. The arrow indicates increasing $\alpha$.}
\label{fig.theta}
\end{figure}

\begin{figure}[top]
\centerline{\includegraphics[scale=0.4]{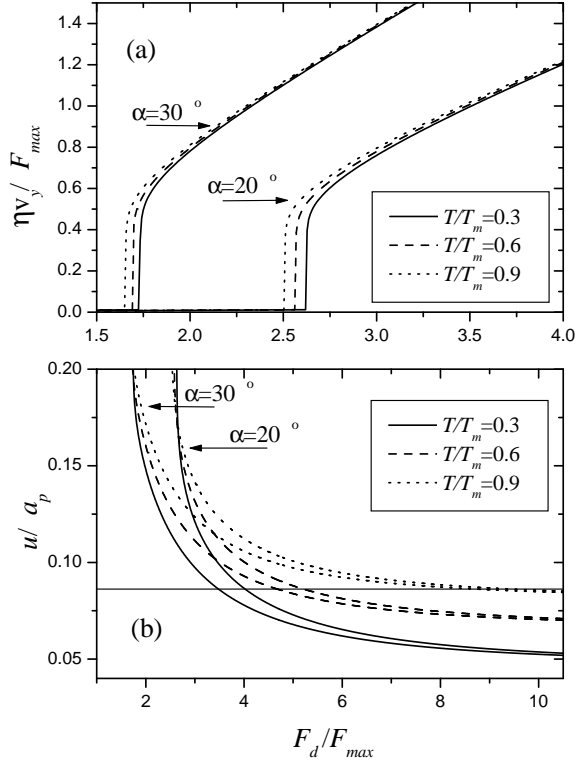}} \vspace{5mm} \caption{
Temperature dependence of: (a) center of mass velocity y-component  and (b)
root-mean- square  displacement curves, $u$ vs. $F_d$.} \label{fig.tdep}
\end{figure}

\begin{figure}[here]
\centerline{\includegraphics[scale=0.4]{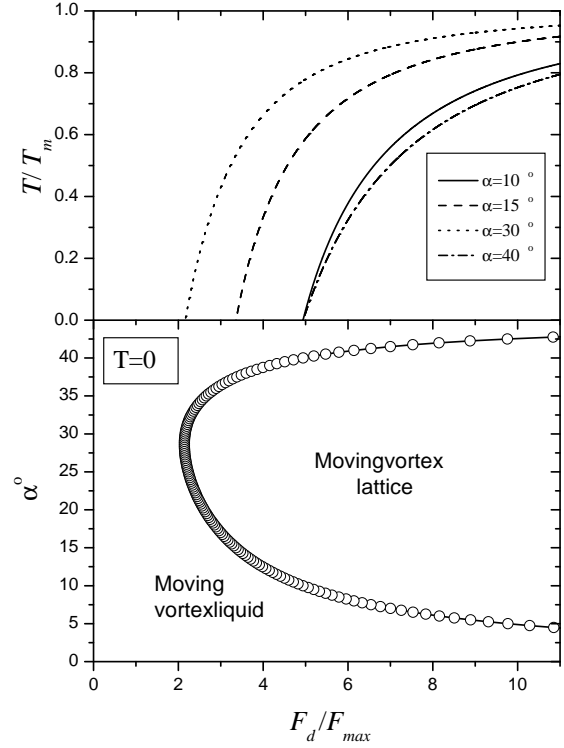}} \vspace{5mm} \caption{Top:
dynamical melting lines for $\alpha$ in the range $10^o\leq \alpha\leq 40^o$.
Bottom: dynamical phase diagram $F_d$ vs. $\alpha$ for $T=0$ .} \label{fig.PD}
\end{figure}

We choose to present the numerical results for the expressions derived in Sec.\
\ref{sec.dyncm}  as functions of the driving force ${\bf F}_d$, instead of
${\bf v}$. To do so we invert Eq. (\ref{eq.vcm3}) by an iterative method. We
start at a high value of the driving force magnitude,  $F_d=50\kappa a_p$, and
a given orientation of ${\bf F}_d$ with respect to the $x$-axis, $\alpha$ (Fig.
\ref{fig.pin}(a)), and approximate the solution  by the result for infinite
drive ${\bf v}={\bf F}_d/\eta$. Then $F_d$ is decreased by small steps, keeping
$\alpha$ fixed. At each step, ${\bf v}$ in the right hand side of Eq.
(\ref{eq.vcm3}) is replaced by the ${\bf v}$ obtained in the previous step.
From the ${\bf v}$ vs. ${\bf F}_d$ relationship thus obtained, the $u$ and
$T_{dm}$ vs. $F_d$ curves follow  using Eqs. (\ref{eq.usq2}) and (\ref{eq.tm}).
The results are shown in Figs.\ \ref{fig.VIu}-\ref{fig.PD}. In the figures
shown here the following units are used. i) Force: $F_{max}=8\pi U_1/a_p=$
magnitude of the maximum force  exerted by the square pinning potential defined
in Eqs. (\ref{eq.pinpot}). ii) Velocities: $F_{max}/\eta$. iii) Temperatures:
$T_m=$ VL equilibrium melting temperature (see Sec.\ \ref{sec.dyncm}).

In Fig.\ \ref{fig.VIu} we show ${\bf v}$ and $u$ as functions of $F_d$ for
several $\alpha$ at $T=0$. The horizontal lines in the bottom panel of Figs.
\ref{fig.VIu}(a) and \ref{fig.VIu}(b) indicate the $u=c_La_v$ lines for a
vortex density corresponding to two vortices per pin ($B=2B_{\phi}$), with
$c_L=0.12$, at which the VL melts, according to Lindemann's criterion.  The
region of  validity of our dynamical cage model is below this line. The $v$ vs.
$F_d$ curves in the top panels of Figs.\ \ref{fig.VIu}(a) and \ref{fig.VIu}(b)
show that the CM velocity is not parallel to the driving force, which leads to
anisotropic V-I curves. This anisotropy is more clearly seen by plotting the
direction of motion, $\theta$, as a function of the driving force magnitude, as
shown in Fig.\ \ref{fig.theta}, for fixed  directions of drive, $\alpha$. These
plots show that, depending on $\alpha$, the moving VL is attracted towards the
[1,0] or [1,1] directions. This behavior is similar to that seen in recent
numerical simulations \cite{gmcr}. Above the $c_L=0.12$ lines in the bottom
panel of Figs. \ref{fig.VIu}(a) and \ref{fig.VIu}(b) we  find a discontinuous
jump in $u$, accompanied by the trapping of moving caged vortex  along [1,0]
for $0<\alpha<32.5^o$ [Fig.\ \ref{fig.VIu}(b)], or along [1,1] for
$32.5^o<\alpha<45^o$ [Fig.\ \ref{fig.VIu}(a)]. This trapping is similar to
transverse pinning observed in numerical simulations and JJA
experiments.\cite{gmc,gmcr,dom1,dom2} This is interesting because it indicates
that, even though our dynamical cage model is not strictly valid in the region
where trapping occurs, it contains physical ingredients capable of describing
the phenomenon. We also note that the jump in $u$ occurs because for vortex
motion along [1,0] and [1,1] there are terms in the sums in Eq.\
(\ref{eq.usq2}) for which ${\bf Q}\cdot{\bf v}=0$. According to the discussion
in Sec.\ \ref{sec.dyncm}, the perturbation expansion for $u$ breaks down and a
large contribution to $u$ results from the terms in the sum  in Eq.\
(\ref{eq.usq2}) with ${\bf Q}\cdot{\bf v}=0$.


The finite-$T$ behavior of the results is shown in Fig.\ \ref{fig.tdep}. We
find that  the ${\bf v}$  vs. $ F_d$ curves change little with $T$ as
illustrated in Fig.\ \ref{fig.tdep}(a). We also find that in  $u^2$ the
$T$-dependent term in  Eq.\ (\ref{eq.usq2}) gives a small contribution.  This
is so because the $T$-dependence in $u$ and ${\bf v}$ results from the third
terms in the right hand side of Eqs. (\ref{eq.usq2})  and (\ref{eq.vcm3}),
which are, essentially, the product of two quantities which are small in the
region of validity of our dynamical cage model.

The dynamical melting lines $T_{dm}$ vs. $F_d$ are shown in the top panel of
Fig.\ \ref{fig.PD} for several $\alpha$. In the bottom panel, the dynamical
phase diagram $F_d$ vs $\alpha$ for $T=0$ is shown.  The $F_d$ vs. $\alpha$
dynamical melting line in the bottom panel corresponds to the intersections of
the $T_{dm}$ vs. $F_d$ curves shown in Fig.\ \ref{fig.PD} top with the $T=0$
line. We find that for other temperatures $T<T_m$ the $F_d$ vs $\alpha$ phase
diagrams are similar.

\section{conclusion}
\label{sec.concl}

In conclusion then, we introduce a simple dynamical model for vortex lattices
flowing elastically  in the presence of periodic pinning. The  model is solved
by a perturbation method that allows the calculation of the dynamical melting
curves,  using Lindemann's criterion,

\noindent and of the V-I curves. This approach is valid for periodic pinning
potentials in general, and predicts that the moving vortex-lattice melting
temperature varies significantly with the CM velocity only for strong pinning.
The perturbation solution is applied to a square pinning lattice. The results
show  qualitative agreement with numerical simulation ones\cite{gmc,gmcr,dom1}.
Namely: i) the predicted dynamical melting temperature approaches the thermal
equilibrium one in the limit of very large center of mass velocities, and
decreases with the velocity. The model predicts that this decrease is,
essentially, as $v^{-2}$. ii) The center of mass velocity is not parallel to
the driving force in general, leading to anisotropic V-I curves. The vortex
lattice is attracted towards the [1,0] or [1,1] direction, depending on the
direction of drive. Transverse pinning, with the vortex lattice motion
restricted to  the [1,0] or [1,1] directions, is found outside the region where
the model is strictly valid.

For other periodic pinning potentials, and for strong pinning, we expect
similar dynamical melting lines and anisotropic V-I curves.

It is significant that the model  predicts anisotropic V-I curves so closely
resembling the numerical simulation ones\cite{gmcr}. These curves follow from
the relationship between the CM velocity and the driving force, Eq.\
(\ref{eq.cag2}). The only approximation used in our solution of this  equation
is the mean-field treatment of Langevin's equations for the whole vortex
lattice. Our results for the V-I curves show that this approximation is
justifiable.

The model can be also applied along the lines described here to study the
effects of weak random potentials, either from  material defects or from
imperfections in the pinning array, on the melting lines and V-I curves. It can
also be generalized to study transverse pinning more rigorously. Work along
these lines is under way and will be reported elsewhere.

\acknowledgments

Research supported in part by CNPq, FAPERJ, FUJB and FACEPE. We thank J. A.
Aguiar and L. R. E. Cabral for stimulating discussions and useful suggestions.

\appendix
\section{}
\label{sec.ap}

To obtain the perturbation expansion of Eq.\ (\ref{eq.cag2}) we write it, using
Eq.\ (\ref{eq.fvpl}), as
\begin{eqnarray}
\eta \frac{d{\bf u}_l(t)}{dt} &=& -\kappa {\bf u}_l(t) + \sum_{\bf Q}{\bf
F}^{\text{\it v-p}}_{\bf Q} e^{i\omega_{\bf Q}t}
\times \nonumber \\
& &\frac{1}{N_v}\sum^{N_v}_{j=1}(N_v \delta_{j,l}-1) e^{i{\bf Q}\cdot [{\bf
R}_j + {\bf u}_j(t)]} + {\bf \Gamma}_l\;, \label{eq.pert1}
\end{eqnarray}
where ${\bf F}^{\text{\it v-p}}_{\bf Q}\equiv -i{\bf Q} U^{\text{\it v-p}}
_{\bf Q}$ and $\omega_{\bf Q}={\bf Q}\cdot{\bf v}$. Expanding $e^{i{\bf Q}\cdot
{\bf u}_j}$ to first order, and carrying out the Fourier transformation in
time, assuming periodic boundary conditions in the time interval $(0,\tau)$,
justifiable for large $\tau$ ($\tau \gg a_p/v$), we find
\begin{eqnarray}
{\bf u}_l(\omega)&=& {\bf u}_l^{T}(\omega)+ \sum_{\bf Q}\frac{{\bf
F}^{\text{\it v-p}}_{\bf Q}}{-i\eta\omega +\kappa}
\frac{1}{N_v}\sum^{N_v}_{j=1} (N_v \delta_{j,l}-1)
\times \nonumber \\
& & e^{i{\bf Q}\cdot{\bf R}_j} \big[\delta_{\omega,-\omega_{\bf Q}}+ i{\bf
Q}\cdot{\bf u}_j(\omega + \omega_{\bf Q})\big] , \label{eq.pert2}
\end{eqnarray}
where ${\bf u}^{T}_l(\omega)$ is the thermal displacement
\begin{equation}
{\bf u}_l^{T}(\omega)=\frac{\Gamma_l(\omega)}{-i\eta\omega +\kappa}\; ,
\label{eq.utom}
\end{equation}
${\Gamma_l(\omega)}$ is the Fourier transform of the random force and
$\delta_K$ denotes the Kronecker delta. The first order solution for ${\bf
u}_l(\omega)$ is obtained by neglecting the linear term in Eq.\
(\ref{eq.pert2}). We find
\begin{equation}
{\bf u}^{(1)}_l(\omega)= {\bf u}^{T}_l(\omega)+\sum_{\bf Q} \frac{{\bf
F}^{\text{\it v-p}}_{\bf Q} e^{i{\bf Q}\cdot{\bf R}_l}} {i\eta\omega_{\bf Q}
+\kappa} \delta_{\omega,-\omega_{\bf Q}}\;. \label{eq.u1}
\end{equation}
To derive Eq.\ (\ref{eq.u1}) we use $\sum^{N_v}_{j=1} e^{i{\bf Q}\cdot {\bf
R}_j}=\delta_{{\bf Q},0}$, since the vortex and pinning lattices are
incommensurate.

The second order solution is obtained by substituting ${\bf u}_j$ in the linear
term in the left hand side of   Eq.\ (\ref{eq.pert2}) by the first order
solution. The result is
\begin{eqnarray}
{\bf u}^{(2)}_l(\omega) &=& {\bf u}^{(1)}_l(\omega)+\sum_{\bf Q}\, \frac{{\bf
F}^{\text{\it v-p}}_{\bf Q}}{-i\eta\omega +\kappa}\, \times
\nonumber \\
& & \frac{1}{N_v}\sum^{N_v}_{j=1}(N_v \delta_{j,l} -1) e^{i{\bf Q}\cdot{\bf
R}_j} i{\bf Q}\cdot{\bf u}^{T}_j(\omega + \omega_{\bf Q}) +
\nonumber \\
& & \sum_{{\bf Q},{\bf Q}^\prime}\frac{{\bf F}^{\text{\it v-p}}_{\bf Q} i{\bf
Q}\cdot{\bf F}^{\text{\it v-p}}_{{\bf Q}^{\prime}} [e^{i{\bf K}\cdot{\bf R}_l}
- \delta_{{\bf Q},-{\bf Q}^\prime}]} {(-i\eta\omega +\kappa)(i\eta\omega_{{\bf
Q}^\prime} +\kappa)} \delta_{\omega,-\omega_{\bf K}}\: ,
\nonumber \\
\label{eq.u2}
\end{eqnarray}
where $\omega_{{\bf Q}^\prime}={\bf Q}^\prime\cdot{\bf v}$, $\omega_{\bf
K}={\bf K}\cdot{\bf v}$, and ${\bf K}={\bf Q}+{\bf Q}^\prime$. With this
solution, one can calculate the mean square displacement for a particular
vortex $l$, $u_l^2=\frac{1}{\tau}\int_0^{\tau}\langle|{\bf u}_l(t)|^2\rangle$,
which up to the second order in the pinning potential is expressed by:
\begin{eqnarray}
u_l^2 &=& \frac{2k_BT}{\kappa} + \sum_{\bf Q,Q^\prime} \frac{{\bf F}^{\text{\it
v-p}}_{\bf Q} \cdot {\bf F}^{\text{\it v-p}}_{{\bf Q}^{\prime}} e^{i{\bf
K}\cdot{\bf R}_l}}{\eta^2\omega_{\bf Q}^2+\kappa^2} \delta_{\omega_{\bf
Q},-\omega_{{\bf Q}^\prime}} +
\nonumber \\
& & \frac{2k_BT}{\kappa}\frac{1}{N_v} \sum_{\bf Q}\frac{Q^2 {\bf F}^{\text{\it
v-p}}_{\bf Q} \cdot {\bf F}^{\text{\it v-p}}_{{\bf Q}^\prime}}
{\eta^2\omega_{\bf Q}^2+4\kappa^2} \times
\nonumber \\
& & \big[(N_v-2)e^{i{\bf K}\cdot{\bf R}_l} + \delta_{{\bf Q},-{\bf
Q}^\prime}\big]\: . \label{eq.usql}
\end{eqnarray}
Taking the average over the $N_v$ vortices and assuming $N_v \gg 1$, Eq.
(\ref{eq.usq2}) is derived.

A similar procedure is used to calculate the driving force dependence of ${\bf
v}$. Expanding Eq. (\ref{eq.vcm}) in the form
\begin{eqnarray}
\eta{\bf v} &=& {\bf F}_d +\frac{1}{N_v}\sum^{N_v}_{j=1}\frac{1}{\tau}
\times \nonumber \\
& & \int^{\tau}_0 \sum_{\bf Q}\big\langle{\bf F}^{\text{\it v-p}}_{\bf Q}
e^{i{\bf Q}\cdot {\bf R}_j}e^{i\omega_{\bf Q}t} \big[1+{\bf Q}\cdot{\bf
u}_j(t)\big]\big\rangle \;, \label{eq.vcm1}
\end{eqnarray}
and using the second order solution for $u_l(t)$ we obtain Eq.\
(\ref{eq.vcm3}).


\end{multicols}
\end{document}